\documentstyle[12pt]{article}
\begin{document}
\title{An Effective Theory for
Midgap States in Doped Spin Ladder and Spin Peierls Systems:
Liouville Quantum Mechanics}
\author{David G. Shelton$^*$ \and Alexei M. Tsvelik$^{\dagger}$}
\date{}
\maketitle
\begin{verse}
* {\it D\'epartement de  Physique et Centre de Recherche en
Physique}\\
{\it  du Solide,
Universit\'e de Sherbrooke, Sherbrooke, Qu\'ebec,}\\
{\it Canada J1K 2R1.} Email: dshelton@physique.usherb.ca \\
$\dagger $ {\it Department of Physics, University of Oxford,
1 Keble Road,}\\
{\it Oxford, OX1 3NP, U.K.} Email: tsvelik@thphys.ox.ac.uk \\
\end{verse}
\begin{abstract}
\par
In gapped spin ladder and spin-Peierls systems the introduction
of disorder, for example by doping, leads to the appearance of
low energy midgap states. The fact that these strongly correlated
systems can be mapped onto one dimensional noninteracting fermions 
provides a rare opportunity to explore  systems which have both
strong interactions and disorder. In this paper we will show that
the statistics of the zero energy midgap wave functions $\psi_0(x)$ 
in these models 
can be effectively described by
Liouville Quantum Mechanics. This enables us to calculate averages
over disorder of the products $\psi_0^2(x_1)\psi_0^2(x_2)...\psi_0^2(x_N)$
(the explicit calculation is performed for $N = 2, 3$). We find
that whilst these midgap states are typically weakly correlated,
their disorder averaged correlations are power law.
 This discrepancy arises because the correlations are not
self-averaging and averages of the wave functions are dominated by anomalously
strongly correlated configurations; a fact which is not always
appreciated in the literature.  
\end{abstract}
\section{Introduction}
One dimensional quantum spin systems have long fascinated both
theorists and experimentalists; more recently there has been
particular interest in the behaviour of such systems in the
presence of disorder (see eg. \cite{fisher} and references 
therein). There are several reasons for this, but one significant
motivation is the comparatively recent experimental realisation
of doped quasi-one-dimensional spin-Peierls and spin ladder systems
\cite{dopes}. One of the most important general results in these 
materials is the appearance upon doping of magnetic states at energies
well below the spin gap- which dramatically changes their magnetic 
properties (see eg. \cite{aff} and references therein).

 As well as the new results emerging from
 experiment, theoretical work has revealed many unusual features 
in a range of different one dimensional disordered spin models.
Prominent among these results is the occurrence of Griffiths 
phases; here the low energy response is dominated by anomalously
strongly correlated regions of the system. The property  related
to that  is a sharp distinction between {\it typical} and 
{\it disorder 
averaged} correlations; the latter are much stronger and typically
power law at criticality \cite{fisher}. The fact that the
 correlation functions are not ``self-averaging'' means
 that rare strongly correlated
configurations dominate the disorder averaged quantities. Thus the 
disorder averaged correlations can be much stronger than one would
naively guess based on the implicit assumption that ``averaged''
and ``typical'' are the same thing. This is important
because it is these averaged
correlations which will be relevant in experiments.

Our approach to these one dimensional spin systems is to exploit
 the various mappings to fermionic theories. This provides an 
alternative way of thinking about the problem which can be related to 
the extensive body of literature on disordered fermion systems.
The spin sector of both the two chain
 spin ladder \cite{me} and the spin-Peierls system \cite{fab}
can be represented in terms of massive noninteracting fermions.
When these systems are lightly doped their conductivity 
shows variable range hopping behaviour, which means that the
charge carriers are strongly localised and no band forms. Thus
to investigate magnetic properties 
it is reasonable to continue to work with a pure spin model
without fluctuations in the charge sector.
 
 We will argue that
doping with nonmagnetic impurities can be 
represented by a type of randomness in the mass of the 
fermions of the effective theories \cite{me},\cite{fab}.
Specifically the mass has fixed magnitude $m_0$ (the value
of the undoped spin gap) except that it flips between $+m_0$
and $-m_0$ at the locations of impurities such that 
$\langle m\rangle =0$. This nontrivial background admits 
localised low energy states well below the gap in addition
 to the massive bulk modes.

It is useful to comment briefly on some insights from the theory
of electrons in disordered metals, in order to better understand 
the connection with spin theories. One of the 
reasons that this problem is so complex is the existence of 
many different length scales. A great deal is known about the
behaviour of metallic-type extended states which explore the
whole sample, i.e. the limit of small wavefunction amplitudes
$t=|\psi(x)|^2$; this depends only upon the global symmetry of
the ensemble and can be derived from random matrix theory 
(see eg. \cite{efetov} and references therein).

However, at length scales much less than the localisation length
$L_c$ but much larger than the mean free path $l$, strongly localised
states with anomalously high local amplitude are important 
(in the conventional localisation this
situation emerges when disorder comes from scattering on magnetic 
impurities). These
states look far from metallic in the region $l\ll L <L_c$ and 
their correlations are
sensitive to local variations in the potential; this is essentially
the origin of the distinction between {\it typical} and
{\it disorder averaged} correlation functions in this regime \cite{multi}.
This is in sharp contrast to the behaviour of  
metallic states which are equally affected by the random potential
at all points in the sample, and not surprisingly the wavefunction
statistics are no longer simple.  Here we can see a relation 
with the behaviour of 1d disordered spin systems; it is this kind
of state which corresponds to localised low energy spin degrees of
freedom.

Recent work on this problem using a variety of approaches has shown that
the wavefunction statistics in this regime can be described by Liouville
field theory \cite{kmt} and that, for example in the two 
dimensional case relevant
to the quantum Hall effect, one observes such interesting phenomena as
multifractality \cite{multi},\cite{kmt}.

It was pointed out in \cite{comtet} that in a system of one dimensional
fermions with random mass $m$ and $\langle m\rangle=0$, the Lyapunov exponent 
$\gamma(E)\sim 1/(-\ln E)$ tends to zero as $E\rightarrow 0$ and so
for low energies we are always at length scales much less than the
 localisation length $L_c\sim 1/\gamma$.
 Thus it is not at all surprising that
in one dimensional systems of length $L>l$ we see a departure from
the universal features predicted by random matrix theory \cite{efetov}.

In this paper we will study one dimensional Dirac fermions with a
random mass. We show that the correlation  functions of the
 so-called prelocalised zero energy
states can be calculated using as an effective theory Liouville
quantum mechanics. This analysis is rather general and does not
depend upon the specific form of the disorder. In section (3) we
will then describe the application of these results to spin systems;
here the low energy states represent the midgap magnetic states 
mentioned above. The advantage of this approach to the midgap states
is that it enables us to get a clear picture of their behaviour and
specifically their correlations, information which is hard to obtain
using more traditional methods.  

\section{Effective Theory}
Suppose we have a system of 1+1 dimensional non-interacting Dirac 
fermions with a random position dependent mass. The Hamiltonian 
can be written;
\begin{eqnarray}
{\cal H}&=&\int dx \psi^{\dagger}\left(-i\sigma^2{\partial\over\partial x}
+\sigma^1 m(x)\right)\psi\\
\psi(x)&=&\left(
\begin{array}{clcr}
u(x)\\
v(x)
\end{array}\right)\label{direq}
\end{eqnarray}
where $\sigma^1$ and $\sigma^2$ are the standard definitions of the 
Pauli matrices.
This leads to the equations of motion, in component form;
\begin{eqnarray}
\left({d\over dx}+m(x)\right)u(x)&=&Ev(x)\nonumber\\
\left(-{d\over dx}+m(x)\right)v(x)&=&Eu(x)\label{eqmo}
\end{eqnarray}
Even if $m(x)$ is finite and 
nonzero almost everywhere, if it passes through zero at
some points there will be normaliseable
 bound states with energy close to zero-  midgap states.
This can be seen from replica, supersymmetry and numerical calculations where 
a peak appears in the density of states at $E=0$
(see eg. \cite{comtet} and references therein). However, within these 
approaches it is notoriously difficult to calculate the correlation 
functions.

It is remarkable that one of the low energy solutions of eqs.(\ref{eqmo}) 
 can be explicitly
written down; it is the zero energy eigenmode;
\begin{eqnarray}
\psi_0(x)&=&{1\over{\cal N}}\left(
\begin{array}{clcr}
1\\
0
\end{array}
\right)\exp\left(-V(x)\right)\label{typical}\\
V(x)&=&\int_0^xm(y)dy\label{v}\\
{\cal N}^{2}&=&\int_0^L dx \exp\left(-2V(x)\right)
\end{eqnarray}
Here it is our purpose to calculate correlations of these zero energy 
eigenstates for the case where $\langle m\rangle$, the average mass is
zero.

In disordered systems, we are usually interested in {\it disorder averaged}
correlations. Here as always the difficulty is to correctly account for
the normalisation factor (denominator) in the average. Following Kogan,
Mudry and Tsvelik \cite{kmt} we introduce a dummy variable $\mu$ which
allows us to transfer all the $V(x)$ dependence from the denominator
into the numerator; a disorder average can then be performed.

In the continuum limit, it is apparent that the quantity $V(x)$  
(\ref{v}) will
behave like the position variable in a random walk \cite{i+d},
where $x$ is the ``time'' coordinate. Therefore, going to 
a path integral formulation we find in the continuum limit
the following Gaussian probability distribution for $V$ 
(for more details see \cite{i+d});
\begin{eqnarray}
P(\{V\}) DV&=&\exp\left[-{1\over 2g}\int dx(\partial_xV)^2\right]DV\label{dist}
\end{eqnarray}
where $g$ parametrises the strength of the disorder.
This will be valid for sufficiently long distances in the correlation
functions; different realisations of disorder may give rise to different
behaviours at small separations. In the case of ``telegraph disorder'' 
where $m$ randomly assumes either $\pm m_0$, $g= 
{\hbar^2m_0^2a_0\over v^2}$ where $a_0$ is the lattice spacing and $v$ is the 
velocity in the Dirac equation (\ref{direq}) (often taken as $1$ for 
simplicity). For more general forms of disorder, $g$ will roughly speaking
be larger for broader distributions of $m$. Note that $g$ is independent
of impurity density- this will however affect the length scales above which
the distribution (\ref{dist}) is valid.

Then the disorder averaged correlator ($\langle X\rangle$ denotes
the disorder average of the quantity $X$) ;
\begin{eqnarray}
\langle|\psi_0(x_1)|^2...|\psi_0(x_N)|^2\rangle
&=&\int DVP(\{V\})|\psi_0(x_1)|^2.....|\psi_0(x_N)|^2
\end{eqnarray}
can be expressed as
\begin{eqnarray}
\langle|\psi_0(x_1)|^2...|\psi_0(x_N)|^2\rangle
&=&\int_0^{\infty}{d\mu\exp[-\alpha\mu]\mu^{N-1}
\over (N-1)!}
\times \int DV\prod_{i=1}^N\mbox{e}^{-2V(x_i)}\mbox{e}^{-S_{\mu}}\nonumber\\
\end{eqnarray}
where $\alpha$ is introduced to regularise the $\mu$ integral 
and
\begin{eqnarray}
S_{\mu}&=&\int dx\left[
{1\over 2g}(\partial_xV)^2+\mu\exp(-2V)\right]
\end{eqnarray}
is the action of Liouville quantum mechanics \cite{hoker} where $x$ is
the ``time'' coordinate.

We can now go to the canonical Hamiltonian form;
\begin{eqnarray}
{\cal H}&=&-{g\over 2}{d^2\over dV^2}+\mu\exp(-2V)
\end{eqnarray}
and find the eigenstates of the 
effective theory, which allows us to calculate the correlation
functions. The detailed calculation of the one,two and three point 
correlators is described in the Appendix. In fact all of the 
correlators are proportional to a constant factor which depends
upon an integral over the dummy variable $\mu $ (which must be
suitably regularised) and an unspecified normalisation factor $A$.
Since the same factor appears in front of all the correlators, we
may fix its value by insisting that the wavefunctions are 
correctly normalised:
\begin{eqnarray}
\int dx \langle|\psi_0(x)|^2\rangle &=&1
\end{eqnarray}
hence
\begin{eqnarray}
\langle |\psi_0(0)|^2\rangle &=&{1\over L}
\end{eqnarray}
where $L$ is the size of the system. Comparing this with the 
result of A2, Eq. (\ref{fixer}) we find for the two and three
point correlators the following correctly normalised power law
behaviour:
\begin{eqnarray}
\langle|\psi_0(x_2)|^2|\psi_0(x_2)|^2\rangle &= &
\left( {1\over L}{g\over 16\sqrt{2\pi}}\right)
\left({1\over gx_{21}}\right)^{3\over 2}\\
\langle|\psi_0(x_3)|^2|\psi_0(x_2)|^2|\psi_0(x_1)|^2\rangle&=&
\left( {1\over L}{g^2\over 1024}\right)
\left({1\over gx_{32}}\right)^{3\over 2}
\left({1\over gx_{21}}\right)^{3\over 2}
\end{eqnarray}
where we have chosen without loss of generality $x_3>x_2>x_1$
and $x_{ij}=x_i-x_j$, and the derivation assumed that 
$|gx_{ij}|\gg 1$, $i\neq j$.
This is a very clear example of the important difference between 
{\it typical} and {\it disorder averaged} correlations.  
From the expression (\ref{typical}) we would expect the typical
correlations to decay exponentially. What happens
 is that atypical strongly correlated configurations
dominate the average and give much stronger power law
correlations in the disorder averaged quantities.

\section{Application to Spin Systems}

We discussed in the introduction some of the characteristic 
features of one dimensional disordered spin systems including
the occurrence of Griffiths phases, and the related distinction
between typical and averaged correlation functions.
As we have seen a similar picture holds for the 
correlations of low energy fermion modes in one dimension; and 
this is not a coincidence but an expression of the deep
relationship  between certain one dimensional spin systems and
systems of fermions.   

Since the analysis above is equally valid for Majorana (real) fermions,
one can first of all make the direct observation that this model with 
random mass is equivalent to the quantum Ising model with random bonds.
This has been studied by many authors (including eg \cite{fisher}, \cite{sm}).
In particular Shankar and Murthy \cite{sm} found that the typical 
correlation functions of fermion bilinears decayed exponentially 
as the square root of distance as one would expect; 
our result shows in addition
that at criticality
the disorder averaged correlation functions
of zero energy modes are power law.

Let us now consider the doped spin-Peierls and spin ladder systems.
Fabrizio and M\'elin \cite{fab} have suggested that one can gain insight 
into the spin-Peierls system by considering the XY version- this maps
to to a system of noninteracting massive Dirac fermions whose mass is
proportional to the dimerisation. They argue that the introduction of
impurities can be modelled simply by the presence of domain walls
between vacua with dimerisation of opposite sign. This leads to a 
model of fermions with a mass which flips between $+m_0$ and $-m_0$ 
at impurities
(by symmetry $\langle m\rangle=0$);
 one finds low energy midgap states localised in regions
with impurities, of which the zero energy 
eigenstate considered in this paper is one example.

 In the language of spins these midgap states correspond to the 
effective spins which appear at breaks or discontinuities in gapped 
spin chains \cite{aff}, since for this model the $z$ component of
the magnetisation $S^z$ is proportional to the fermion density.

 These impurity spins
 are often treated in the literature as
though they are essentially uncorrelated, based on the argument
that they are {\it typically} weakly coupled for low impurity 
density. We emphasise again that whilst this is true, our work
shows that their disorder averaged correlations are power law;
and these are the relevant quantities in experiment. It is for
this reason, for example that they do not give a simple Curie contribution
to the susceptibility; Fabrizio and M\'elin  \cite{fab} showed that
$\chi (T)\sim 1/(T\ln^2T)$ a result which differs significantly 
at low temperatures from the noninteracting $1/T$ result.

Of course in reality
the problem of spin-Peierls systems is more complicated-
there are very important three dimensional effects and thermal
fluctuations of the lattice distortion.
It is also unclear, even in an inherently one dimensional system,
whether the introduction of impurities can really be represented 
so simply \cite{fab}.
 Nonetheless we would argue
that we have captured a crucial feature; which is that at boundaries
between the two degenerate ground states (domain walls $\equiv$ kinks
in the fermion mass) we find localised magnetic low energy 
degrees of freedom. 
 
Similar arguments applied to a fermionic model of the two chain 
spin ladder \cite{me}, where doping is introduced in the form of
 static kinks
in the charge field, leads again to a model of fermions with mass
flipping between $-m_0$ and $m_0$ at the location of impurities.
This can be seen from the bosonisation formalism, treating 
nonmagnetic impurities in a way suggested by Fukuyama {\it et al.}
\cite{fuku}. At half filling in a spin ladder system, the charge 
modes will be frozen (gapped). Nonetheless, because the charge density
is proportional to the gradient of the charge field, a nonmagnetic
charged impurity (hole) can be represented as a static kink in the
field. This assumes that there are sufficiently few impurities and
that they are sufficiently strongly localised so that no band forms-
these conditions appear to be satisfied in typical experimental 
systems \cite{dopes}. 

By taking these kinks into account in the effective theory of
\cite{me}, it can be seen that their effect is to introduce
kinks in the mass of the fermions describing the spin sector
\cite{nersesyan}. 

The interpretation in terms of spins is much as for the spin-Peierls
system. In the effective theory \cite{me}, the slow component of the
magnetisation is essentially proportional to the fermion density, and 
so these localised states can again be interpreted as low energy spin
degrees of freedom.
 
\section{Conclusions}
We have shown that Liouville quantum mechanics can usefully be 
applied to calculate the properties of the zero energy localised
states in a model of Dirac fermions with random mass. We have emphasised 
the relation of this model to some one dimensional disordered spin systems;
one of the most important points is that at the critical point ($\langle
m\rangle =0$)
we find power law correlations
for the disorder averaged correlation functions, even though typical 
correlations decay exponentially. This shows that even if midgap states
are typically weakly correlated, anomalously strongly correlated 
configurations can dominate in physical quantities.

Finally, it is very interesting in itself that one can derive an
effective theory for midgap states with nontrivial, calculable
correlation functions; although a
limitation of our approach is that it only gives information about
states with energy $E=0$. It would be very interesting if the 
current approach could be extended to investigate
states with finite energy.
\section{Acknowledgements}
 A. M. T. is  grateful to I. I. Kogan, A. A. Nersesyan, Yu Lu and M. Fabrizio
for fruitful discussions and interest in the work. D.G.S. thanks D. 
S\'en\'echal
for helpful comments and discussions and A.A. Nersesyan for drawing
his attention to these systems.
 
\appendix
\section{Calculation of Correlators}
\subsection{Two Point Correlator}
\begin{eqnarray}
C_{21}&=&\langle |\psi_0(x_2)|^2|\psi_0(x_1)|^2\rangle\\
C_{21}&=&\int \mu d\mu \mbox{e}^{-\alpha\mu}M_2\\
M_2&=&\int_{-\infty}^{+\infty}\int_{-\infty}^{+\infty}  dV_1dV_2 \langle 0|_2
\mbox{e}^{-2V_2(x_2)}|u\rangle_2\langle u|_1
\mbox{e}^{-2V_1(x_1)}|0\rangle_1\\
&=&\int_0^{+\infty}dE \left|\int_{-\infty}^{+\infty}
 dV\langle 0|\mbox{e}^{-2V(0)}|u\rangle\right|^2\mbox{e}^{-Ex_{21}}
\end{eqnarray}
where $E=gu^2/2$,  we have chosen without loss of 
generality $x_{21}=x_2-x_1>0$ and the 
energy normalised eigenstates of Liouville quantum mechanics 
at $x=0$ are 
given as follows \cite{hoker};
\begin{eqnarray}
|0\rangle &=&A K_0\left(\sqrt{2\mu\over g}\mbox{e}^{-V}\right)\\
|u\rangle &=&\left[{1\over 2\pi g}\sinh\pi u\right]^{1\over 2}
K_{iu}\left(\sqrt{2\mu\over g}\mbox{e}^{-V}\right)
\end{eqnarray}
where A is a normalisation which we can fix later.
The ``time'' ($x$) evolution of these states is denoted by
subscripts on the bras and kets:
\begin{eqnarray}
|u\rangle_i&=&\mbox{e}^{-Ex_i}|u(V=V_i)\rangle
\end{eqnarray}
First let us work out the integral over $V$, making the substitution
$x=(\sqrt{2\mu/g})\mbox{e}^{-V}$;
\begin{eqnarray}
I&=&\int_{-\infty}^{+\infty} dV 
\langle 0|\mbox{e}^{-V}|u\rangle\label{mel}\\
I&=&{Ag^{1\over 2}\over 2\pi\mu}\left[\sinh\pi u\right]^{1\over 2}
\int_0^{+\infty} xK_0(x)K_{iu}(x)dx
\end{eqnarray}
Which gives \cite{gnr};
\begin{eqnarray}
I&=&{Ag^{1\over 2}\over 8\pi\mu}\left[\sinh\pi u\right]^{1\over 2}
\left({{\pi u\over 2}\over \sinh{\pi u\over 2}}\right)^2  
\end{eqnarray}
So now 
\begin{eqnarray}
M_2&=&{A^2g^2\over 8\pi\mu^2}\int_0^{+\infty} udu\sinh\pi u
\left({{\pi u\over 2 }\over \sinh {\pi u\over 2}}\right)^4
\exp(-{gx_{21}\over 2}u^2)
\end{eqnarray}
For long distances $gx_{21}\gg 1$ the Gaussian factor in the integrand is
very rapidly decaying, so the other parts of the integrand can be expanded
for small $u$, giving;
\begin{eqnarray}
M_{2}&\approx &{A^2g^2\over 64\pi\mu^2}\int_{0}^{\infty}
u^2\exp(-{gx_{12}\over 2}u^2)du\\
&\approx &{A^2g^2\over 64\sqrt{2\pi}\mu^2}\left({1\over gx_{21}}
\right)^{3\over 2}
\end{eqnarray}
which gives for the two point correlation function at long distances;
\begin{eqnarray}
C_{21}&\approx &{A^2g^{1\over 2}\over 64\sqrt{2\pi}}\left[\int_0^{+\infty}{d\mu\over\mu}
\mbox{e}^{-\alpha\mu}\right]\left({1\over x_{21}}\right)^{3\over 2}
\end{eqnarray}
\subsection{Normalisation}
To set the normalisation let us consider
\begin{eqnarray}
C_{1}&=&\langle |\psi_0(0)|^2\rangle\\
&=&\int d\mu\mbox{e}^{-\alpha\mu}M_1
\end{eqnarray}
where 
\begin{eqnarray}
M_1&=&\int_{-\infty}^{+\infty}
dV\langle 0|\mbox{e}^{-2V}| 0\rangle\nonumber\\
&=&A^2\int_{-\infty}^{+\infty}dV
\left[K_0\left(\sqrt{2\mu\over g}\mbox{e}^{-V}\right)\right]^2
\mbox{e}^{-2V}
\end{eqnarray}
Again making the substitution $x=(\sqrt{2\mu/g})\mbox{e}^{-V}$ we
find that 
\begin{eqnarray}
\langle|\psi_0(0)|^2\rangle &=&{A^2g\over 4}\left[
\int_0^{+\infty}{d\mu\over\mu}\mbox{e}^{-\alpha\mu}\right]\label{fixer}
\end{eqnarray}
\subsection{Three Point Correlator}
\begin{eqnarray}
C_{321}&=&\langle |\psi_0(x_3)|^2|\psi_0(x_2)|^2|\psi_0(x_1)|^2\rangle\\
C_{321}&=&\int_0^{+\infty}d\mu{\mbox{e}^{-\alpha\mu}\mu^2\over 2}M_3
\end{eqnarray}
Inserting the resolution of unity as before;
\begin{eqnarray}
M_3&=&\int_0^{+\infty}\int_0^{+\infty}dE_1dE_2
(I_1I_2I_3)\label{int}\\
I_1&=&\int_{-\infty}^{+\infty} dV_1\langle 0|\mbox{e}^{-2V_1(0)}|u_1\rangle
\mbox{e}^{-E_1x_1}\\
I_2&=&\int_{-\infty}^{+\infty} dV_2\langle u_1|\mbox{e}^{-2V_2(0)}
|u_2\rangle \mbox{e}^{(E_1-E_2)x_2}\\
I_3&=&\int_{-\infty}^{+\infty}
dV_3\langle u_2|\mbox{e}^{-2V_3(0)}|0\rangle\mbox{e}^{E_2x_3}
\end{eqnarray}
where $E_i=(g/2)u_i^2$, $i=1,2$. The integrals $I_1$ and $I_3$ are the same
as eq.
(\ref{mel})whereas $I_2$ is a little different- introducing again the variable
$x=(\sqrt{2\mu/g})\mbox{e}^{-V}$;
\begin{eqnarray}
I_2&=&{1\over 4\pi\mu}\mbox{e}^{(E_1-E_2)x_2}
(\sinh\pi u_1\sinh\pi u_2)^{1\over 2}
\int_{0}^{+\infty}xdxK_{iu_1}(x)K_{iu_2}(x)\\
&=&{1\pi\over 32 \mu}\mbox{e}^{(E_1-E_2)x_2}
(\sinh\pi u_1\sinh\pi u_2)^{1\over 2}{u_1^2-u_2^2\over \cosh u_1-\cosh u_2}
\end{eqnarray}
Choosing without loss of generality $x_{21}=x_2-x_1>0$ and
$x_{32}=x_3-x_2>0$ we can use the same argument as before, that
for large separations the exponentials of energy will decay very
rapidly, and so the other parts of the integrand 
(\ref{int}) can be expanded for
small $u_1,u_2$. Then we find;
\begin{eqnarray}
M_3&\approx &{A^2g^3\over 1024\pi\mu^3}
\int_0^{+\infty}u_1^2du_1\exp(-{gx_{21}u_1^2\over 2})
\int_0^{+\infty}u_2^2du_2\exp(-{gx_{32}u_2^2\over 2})
\end{eqnarray}
And we obtain for the 3 point correlation function at long distances;
\begin{eqnarray}
C_{321}&=&
{A^2\over 4096}\left[\int_0^{+\infty}
{d\mu\over\mu}\mbox{e}^{-\alpha\mu}\right]
\left({1\over x_{21}}\right)^{3\over 2}
\left({1\over x_{32}}\right)^{3\over 2}
\end{eqnarray}

 \end{document}